\documentclass{amsart}
\usepackage[latin1, utf8]{inputenc}

\usepackage{amsmath,amsfonts,amssymb,amsthm}
\usepackage{physics}
\usepackage{subfig}
\usepackage{graphicx}
\usepackage{appendix}
\usepackage{comment}
\usepackage{todonotes}
\usepackage[foot]{amsaddr}

\DeclareMathOperator*{\argmin}{arg\,min}

\newtheorem{definition}{Definition}
\newtheorem{theorem}{Theorem}

\title{Evaluating the Convergence Limit of Quantum Neural Tangent Kernel}
\author{Trong Duong}
\address{Korea Advanced Institute of Science and Technology (KAIST)}
\thanks{This work was conducted during my internship at the Centre for Quantum Technologies (CQT), Singapore, under the valuable guidance of Patrick Rebentrost.}
\date{July 2022}

\usepackage{biblatex}
\begin{document}

\maketitle

\begin{abstract}
    Quantum variational algorithms have been one of major applications of quantum computing with current quantum devices. There are recent attempts to establish the foundation for these algorithms. A possible approach is to characterize the training dynamics with quantum neural tangent kernel. In this work, we construct the kernel for two models, Quantun Ensemble and Quantum Neural Network, and show the convergence of these models in the limit of infinitely many qubits. We also show applications of the kernel limit in regression tasks.
\end{abstract}

\section{Introduction}
The study of overparametrized models in machine learning has drawn attentions thanks to the increasing interest in deep learning. These models, which have many more parameters than training samples, have empirically good learning capability and performance on the test data. This unexpected generalization may be explained by a training algorithm such as gradient descent somehow inducing implicit regularization of the model. In particular, prior work showed that in some circumstances, even complex nonlinear models might behave as if in the kernel-based training with regularization on the corresponding Reproducing Kernel Hilbert Space (RKHS) \cite{jacot2018neural}. In contrast, other work showed that under different conditions some models can exhibit different non-Hilbert norm implicit regularization \cite{gunasekar2017implicit, gunasekar2018implicit, savarese2019infinite, woodworth2020kernel}.

On the other hand, quantum variational models is a common class of algorithms in the noisy intermediate-scale quantum (NISQ) era. Essentially, the model uses the natural quantum objective function $f(x,\theta)) = \bra{\psi} H \ket{\psi}$ for some Hermitian observable $H$, where $\ket{\psi} = U(x,\theta)\ket{0}$ is prepared by applying unitary identified by the input and model's parameters on the initial state. The unitary matrix contain information about parameters and input data. There are prior work devoted to establish the notion and convergence of quantum neural tangent kernel \cite{shirai2021quantum, liu2022representation}, prove the fair constant-ness of the tangent kernel at initialization and during training \cite{liu2023analytic, abedi2023quantum}, and the effect of noisy measurements on the kernel \cite{liu2022laziness}.

We have so far a certain number of quantum variational models that could entail a converging tangent kernel. The limit of the kernel at the limit of infinite dimension is particularly useful. For example, there are efficient analytical form of the tangent kernel for classical fully connected neural networks and convolutional neural networks at the infinite limit \cite{jacot2018neural, arora2019exact}, which serve as an effective benchmark for models of the same kind. From my humble experience, there has not yet an analytical form for quantum neural tangent kernel. This is because quantum neural networks are usually realized by some specific architecture the convergence limit of the kernel depends on. Hence to utilize the benefits of the kernel in high dimensional space, one still have to carry out operations of large unitary matrices. The computation quickly becomes inefficient as the dimension of the Hilbert space scales exponentially to the number of qubits.

We attempt to provide a quantum model where one can efficiently compute the analytical neural tangent kernel at large dimension  without perform large matrix operations.

\section{Preliminaries}
\subsection{Neural Tangent Kernel}
It has been shown that training a neural network with gradient descent is highly related to ``neural tangent kernel''. In particular, when the neural network can be approximated by a linearized model, one can describe the dynamics of the model by a simple first-order differential equation. The situation happens in the so-called ``kernel regime'', which usually occurs in an overparametrized model in which every parameter stays almost unchanged \cite{jacot2018neural}. We first specify the relations between training a neural network and  the neural tangent kernel and conditions for the kernel regime to arise.

\paragraph{Neural Tangent Kernel}
Let the loss function to be optimize is $L(\theta) = \frac{1}{2} \sum_i |f(\mathbf{x}_i;\theta) - y_i|^2 = \frac{1}{2}\|\mathbf{f}(\mathbf{X},\theta) - \mathbf{y}\|^2$. Parameters are updated at every step according to gradient descent $\delta \theta = \theta(t+1) - \theta(t) = -\eta \nabla L(\theta(t))$. With infinitesimal learning rate $\eta$, the update rule is equivalent to the gradient flow
\begin{align}
    \dot{\theta} = \frac{\partial \theta}{\partial t} &= \nabla L(\theta(t)) \\
    &= - \nabla \mathbf{f}(\theta) (\mathbf{f}(\theta) - \mathbf{y})
\end{align}
where $\mathbf{f}(\theta) \equiv f(\mathbf{X};\theta)$. We can induce the dynamics of the model in the function space 
\begin{align}
    \dot{\mathbf{f}}(\theta) &= \nabla \mathbf{f}(\theta)^T \dot{\theta} \\
    &= - \nabla \mathbf{f}(\theta)^T \nabla \mathbf{f}(\theta) (\mathbf{f}(\theta) - \mathbf{y}) \\
    &= - \mathbf{K}(\theta) (\mathbf{f}(\theta) - \mathbf{y})
\end{align}
The quantity $\mathbf{K}(\theta) = \nabla \mathbf{f}(\theta)^T \nabla \mathbf{f}(\theta)$ is called the ``neural tangent kernel'' (NTK). In some special cases when this kernel stay almost constant during training $\mathbf{K}(\theta(t)) \approx \mathbf{K}(\theta_0)$ the training dynamics can be described by linear first-order equation $\dot{\mathbf{f}}(\theta) = - \mathbf{K}(\theta_0) (\mathbf{f}(\theta) - \mathbf{y})$, which turns out to have a simple solution
\begin{align}
    \mathbf{f}(\theta(t)) = \mathbf{y} + e^{- \mathbf{K}(\theta_0) t} (\mathbf{f}(\theta_0)) - \mathbf{y})
\end{align}

The situation when the kernel is constant is referred to as the ``kernel regime'' or ``lazy regime'', usually characterized by constant gradient $\nabla \mathbf{f}(\theta(t)) \approx \nabla \mathbf{f}(\theta_0) \Rightarrow \mathbf{K}(\theta(t)) \approx \mathbf{K}(\theta_0)$.
This happens when the parameters do change much during training, i.e. they stay close to the initialization. Expand the output function near the initialization
\begin{align}
    f(\mathbf{x};\theta) = f(\mathbf{x};\theta_0) + \nabla f(\mathbf{x};\theta_0)^T (\theta - \theta_0)
\end{align}
Notice that the output function $f(\mathbf{x};\theta)$ is linear in the parameters but non-linear on the input data with data feature $\Phi(\mathbf{x}) = \nabla f(\mathbf{x};\theta_0)^T$.

It was proved that the linearization that assumes fairly constant gradient during training occurs when every hidden layer of the neural network has infinite width and parameters initialized with normal distributions \cite{jacot2018neural},
and when the output function is scaled by some big factor $f \mapsto \alpha f$ \cite{chizat2019lazy}. Moreover, for infinitely wide neural networks, the kernel is fairly constant over different initializations $\theta_0$ and there is a recursive formula to calculate the kernel $\mathbf{K}_\infty(\theta_0)$ the kernel converges to at the infinite limit. The proof of convergence for classical neural network heavily relies on the fact that the neural network behaves like a Gaussian process at the infinite-width limit. In this project we attempt to find such a convergence for the neural tangent kernel of a quantum neural network (QNN).

\subsection{Reproducing kernel Hilbert space}
To understand the importance of kernel to (quantum) machine learning, we have to study Reproducing Kernel Hilbert Space (RKHS), the central concept of kernel theory. The type of kernel that is particularly useful is positive definite and symmetric (PDS) kernel, which is almost always implied in the literature about kernel methods. We provide a brief introduction to the kernel theory. Readers can find a comprehensive discussion about the kernel method in \cite{mohri2018foundations}.
\begin{definition}
A kernel is a real-valued bivariate function $k: \mathcal{X} \times \mathcal{X} \rightarrow \mathbb{R}$ defined over the input space $\mathcal{X}$. The kernel is positive-definite and symmetric (PDS) if
\begin{enumerate}
    \item for all $m \in \mathbb{N}, c_i \in \mathbb{R}, x_i \in \mathcal{X}, i \in [m]$,
    \begin{equation}
        \sum_{i,j=1}^m c_i c_j k(x_i,x_j) \geq 0.
    \end{equation}
    
    \item $k(x,x') = k(x',x)$ for all $x,x' \in \mathcal{X}$.
\end{enumerate}
\end{definition}
For example, the tangent kernel, or any kernel defined by an inner product, is PDS because $\sum_{i,j} c_i c_j \nabla f(x_i)^T \nabla f(x_j) = \|\sum_i c_i\nabla f(x_i)\|^2$ and the symmetricity is straightforward. A key property of PDS kernels is that there exist a Hilbert space, i.e, a complete vector space equipped with inner product, called Reproducing Kernel Hilbert Space. This RKHS should be distinguished from the quantum Hilbert space $\mathcal{H}$, hence we denote it by $\mathcal{G}$.
\begin{theorem}[Reproducing Kernel Hilbert Space]
Given a positive-definite and symmetric kernel function $k$ over the input space $\mathcal{X}$, there exists a Hilbert space $\mathcal{G}$ and a map $\phi: \mathcal{X} \rightarrow \mathcal{G}$ such that
    \begin{enumerate}
        \item $\mathcal{G} = \operatorname{span}\{g_x: x \in \mathcal{X} \}$ and shall include the limit of every Cauchy sequence, where $g_x: \mathcal{X} \rightarrow \mathbb{R}$ is a functional given by $g_x(\cdot) = k(x,\cdot)$.
        \item $\phi(x) \in \mathcal{G}$ defined by $\phi(x) = g_x = k(x,\cdot)$
        \item $k(x,x') \equiv \expval{k(x,\cdot), k(x', \cdot)}_{\mathcal{G}} = \expval{\phi(x), \phi(x')}_{\mathcal{G}}$ for all $x,x' \in \mathcal{X}$.
        \item (Reproducing property) For all $h \in \mathcal{G}$,
            \begin{equation}
                h(x) = \expval{h, \phi(x)}_{\mathcal{G}} = \expval{h, k(x,\cdot)}_{\mathcal{G}}
            \end{equation}
    \end{enumerate}
\end{theorem}
The reproducing property comes from the construction of $\mathcal{G}$ that we can write $h = \sum_{i=1} a_i k(x_i, \cdot), x_i \in \mathcal{X}$. Then $\phi(x)$ ``reproduces'' the values $h(x)$ for every $h \in \mathcal{G}$,
\begin{equation}
    h(x) = \sum_{i} a_i k(x_i,x) = \expval{\sum_{i}a_i k(x_i, \cdot), k(x,\cdot)}_{\mathcal{G}} = \expval{h, \phi(x)}_{\mathcal{G}}.
\end{equation}

The application of kernel and RKHS in machine learning comes from the famous Representer Theorem, which establishes the fact that we only need to search through a linear combination of finite kernel functionals for a supervised learning problem.
\begin{theorem}[Representer Theorem]
Given a kernel $k$, its induced RKHS $\mathcal{G}$ and feature map $\phi$. Define an objective function $F: \mathcal{G} \rightarrow \mathcal{R}$ as a regularized empirical loss function wrt. some dataset $\{(x_i,y_i), i\in [m]\}$
\begin{equation}
    F(h) = G(\|h\|_{\mathcal{G}}) + L( (h(x_1),y_1), \dots,(h(x_m),y_m) )
\end{equation}
such that $G(\cdot)$ is non-decreasing. If $\argmin_{h\in \mathcal{G}} F(h)$ has a solution, then there exists a solution of the form $h' = \sum_{i=1}^m a_i k(x_i,\cdot) = \sum_{i=1}^m a_i \phi(x_i)$. Furthermore, if $G$ is strictly increasing, all solutions must have the given form.
\end{theorem}
The problem of minimizing $F(h)$ with mean square loss function, where $h$ admits the form $\sum_{i=1}^m a_i \phi(x_i)$ can be solved by ordinary least square with the kernel trick that essentially utilizes the simple evaluation of $k(x,x')$ in many cases. Moreover, it was shown that variational quantum models, a huge class of quantum machine learning models, are equivalent to functions in the corresponding RKHS \cite{schuld2021supervised}. Hence finding an optimal model can be solved by finding an optimal solution on the RKHS.

\section{Quantum Ensemble Model}
We consider a model that, given a collection $\{(a_n, U_n, W_n): a_n \in \mathbb{R}, U_n, W_n \in \mathcal{U}(d) \}, n \in [N]$, where $\mathcal{U}(d)$ is the unitary group of dimension $d$. For a qubit system of $n$ qubits, the corresponding quantum Hilbert space has the dimension of $d=2^n$. Define a model
\begin{equation}
    \label{eq:ensemble-model}
    f(x) = \frac{\sqrt{d}}{\sqrt{N}} \sum_{n=1}^N a_n \bra{0}U_n^\dagger S^\dagger(x) W_n^\dagger H W_n S(x) U_n \ket{0}.
\end{equation}
This type of model, which looks like an ensemble of natural quantum objective functions, is non-conventional. However it would shed light on the concept of tangent kernel in the context of quantum machine learning. With a training dataset $\{(x_p,y_p): y_p \in \mathbb{R}\}, p \in [P]$, the model can be optimized with respect to $a_n$, which is an ordinary least square problem.
\begin{align}
    & \min_{a} \frac{1}{2} \sum_{p=1}^P |f(x_p)-y_p|^2 \\
    =& \min_{a} \frac{1}{2} ||Fa -y||^2,
\end{align}
where F is a matrix whose elements are given by 
\begin{equation}
    F_{pn} = \frac{\sqrt{d}}{\sqrt{N}} \bra{0}U_n^\dagger S^\dagger(x_p) W_n^\dagger H W_n S(x_p) U_n \ket{0},
\end{equation}
i.e. every element is the expectation of the observable $H$ evaluated at the state $\ket{\psi_{pn}} = W_n S(x_p) U_n \ket{0}$. The solution to this problem is
\begin{equation}
    \hat a = (FF^T)^{-1}Fy
\end{equation}
For the model to have good performance, one has to has a sufficiently large $N$, and the unitaries $U_n, W_n$ should be diverse. A general guide to choose $N$ and the unitaries seems to be lacking up to my knowledge.

On the other hand, we can observe that the gradient $\nabla_a f(x_p)$ is also the $p$-th row of $F$. From the definition of tangent kernel, the kernel matrix is
\begin{equation}
\begin{aligned}
    K(x_p, x_{p'}) &= \nabla_a f(x_p)^T \nabla_x f(x_{p'}) \\
    &= \frac{d}{N} \sum_{n=1}^N \bra{\psi_{pn}} H \ket{\psi_{pn}} \bra{\psi_{p'n}} H \ket{\psi_{p'n}}.
\end{aligned}
\end{equation}
Although this kernel matrix is independent to $a$, it depends on the choice of unitaries. Suppose we have no prior knowledge about selected unitaries and assume them to be sampled independently from $\mathcal{U}(d)$ with the Haar measure. In that case one can derive the expectation value of a polynomial of entries of random unitaries with respect to the unitary Haar measure \cite{puchala2011symbolic, collins2006integration}. Here we use a symbolic integration program using tensor network \cite{fukuda2019rtni} to simplify the computation.

\begin{align}
    \nonumber \mathbb{E}K(x_p,x_{p'})=& d \times \mathbb{E}_{\mathcal{U},\mathcal{W}} \left[ \bra{0}U^\dagger S^\dagger(x_p) W^\dagger H W S(x_p) U \ket{0}  \bra{0}U^\dagger S^\dagger(x_{p'}) W^\dagger H W S(x_{p'}) U \ket{0} \right] \\
    \nonumber =& d \times \frac{d^2\Tr(H)^2 + d\Tr(H)^2 + d\Tr(H^2) + \Tr(H^2)s_{pp'}}{(d^2-1)^2}\\
    \nonumber +& d\times \frac{d^2\Tr(H^2) + d\Tr(H)^2 + d\Tr(H^2) + \Tr(H)^2s_{pp'}}{d^2(d^2-1)^2} \\
    \nonumber -& d\times \frac{d^2\Tr(H)^2 + d^2\Tr(H^2) + 2d\Tr(H)^2 + 2d\Tr(H^2) + (\Tr(H)^2 + \Tr(H^2))s_{pp'}}{d(d^2-1)^2} \\
    =& \frac{d^3+d^2-d-s_{pp'}}{d(d^3+d^2-d-1)}\Tr(H)^2 + \frac{s_{pp'}-1}{d^3+d^2-d-1}\Tr(H^2)
\end{align}
where $s_{pp'} = |\Tr(S(x_{p'}) S^\dagger(x_p))|^2$. A class of common observable operators in practice is Pauli measurement, i.e. $H \in \{I,X,Y,Z\}^n$, all of which satistfy $\Tr(H) = 0$ and $\Tr(H^2) = d$. In the limit $d \rightarrow \infty$, the limit of $\mathbb{E} K(x_p, x_{p'})$ aligns with the intuition that kernel is a similarity measure.
\begin{align}
    \mathbb{E} K(x_p, x_{p'}) & = \frac{d(s_{pp'}-1)}{d^3+d^2-d-1} \\
    &\rightarrow \left\{\begin{matrix}
        1,& \text{ if } s_{pp'} \approx d^2 \\
        0,& \text{ o.w.}
        \end{matrix}\right.
\end{align}
The condition for the positive limit to hold is $S(x_p) \approx S(x_{p'})$. For example, when the encoding block is $S(x) = \left(e^{-ixZ/2}\right)^{\otimes n}$, the kernel looks like a partial Fourier sum as shown in Appendix~\ref{appx: pauliZ-encoding}
\begin{align}
    s_{pp'} = \binom{2n}{n} + 2\sum_{k=1}^n \binom{2n}{n-k} \cos(k(x_p - x_{p'}))
    \label{eq:partial-fourier}
\end{align}
In most of the cases, there are no such efficient analytical forms to compute $s_{pp'}$. Fortunately, one can estimate it with the Hadamard test. Considering a qubit system with a probe qubit initialized at $\ketbra{0}{0}$ and $n$ qubits at the completely mixed state $\frac{1}{d} I_d, d=2^n$. The mixed state can be prepared by applying the Pauli-$X$ gate on each qubit uniformly at random. One can prepare the system into the state
\begin{align}
    \rho_{n+1} = \frac{1}{2d} \left(\ketbra{0}{0} I_d + \ketbra{0}{1}U^\dagger +  \ketbra{1}{0}U + \ketbra{1}{1}I_d\right) = \frac{1}{2d} \begin{pmatrix}
    I_d & U^\dagger\\ 
    U & I_d \end{pmatrix}
\end{align}
by applying a Hadamard gate on the probe qubit, followed by $U = S(x_{p'}) S^\dagger (x_p)$ in the ancillary system controlled on the probe qubit. Measuring the probe qubit gives sufficient information to compute $\Tr(U)$ since $\expval{X} = \frac{1}{d}\Re\Tr(U)$ and $\expval{Y} = -\frac{1}{d}\Im\Tr(U)$. A simple Chernoff bound can show that one needs $O(1/\varepsilon^2)$ samples to estimate either the real or imaginary part within the absolute error $\varepsilon$, independent to the size of the system. Hence if restrained to use only multiple parallel local encoding blocks for the encoding unitary $S(x) = \left(e^{-ix\sigma}\right)^{\otimes n}$, we can evaluate the kernel matrix efficiently regardless of the dimension. This provides a quick alternative technique for regression and classification problems.

\begin{figure}[t] \centering
    \subfloat[Standard techniques]{
    \includegraphics[width=0.4\textwidth]{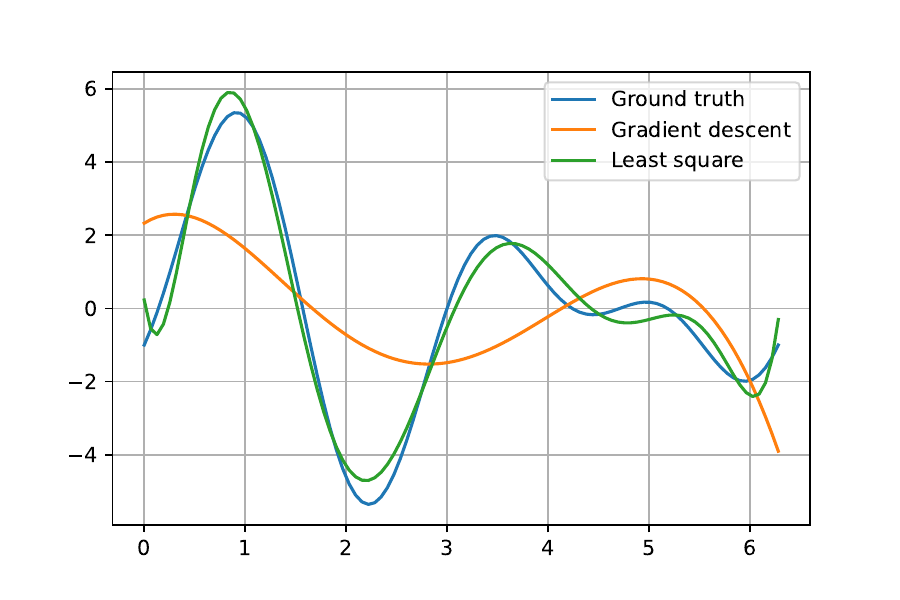}
    }
    \subfloat[Kernel method] {
    \includegraphics[width=0.4\textwidth]{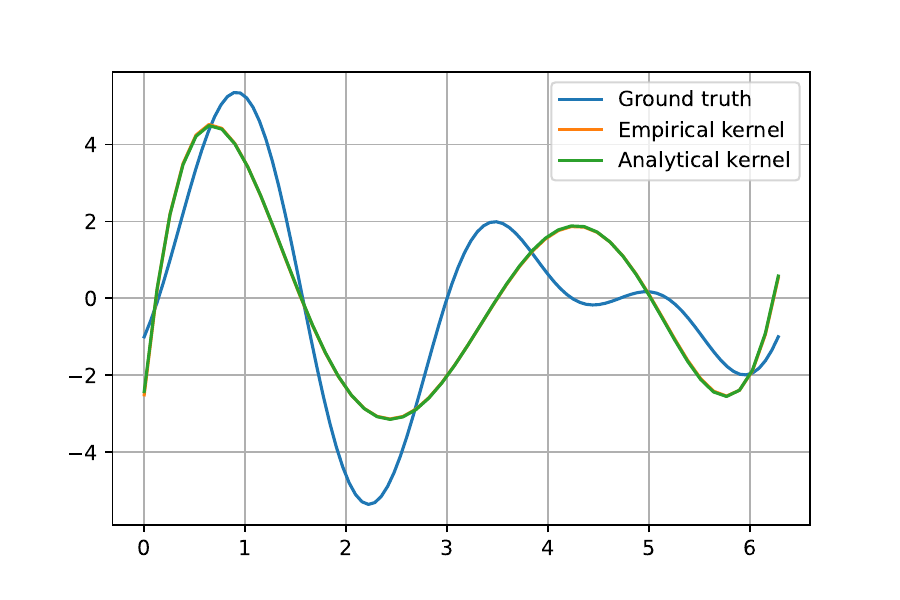}
    }
    \caption{Optimization methods for quantum ensemble model \eqref{eq:ensemble-model} with $n=4$ qubits and $N=2000$ terms. The function to learn is $f(x) = \cos(x)+3\sin(2x)-2\cos(3x)$ defined on $[0,2\pi]$.}
\label{fig:ensemble}
\end{figure}

Although the kernel in the ensemble model might have the form of a partial Fourier sum \eqref{eq:partial-fourier}, the corresponding kernel function fails be universal in the sense that it can approximate any function in the RKHS within arbitrary precision. This is because we have no control over the coefficients of the Fourier sum even though we can have as many Fourier frequencies as desired.

\section{Quantum Neural Network Model}
Define a quantum neural network to have the form $WS(x)U\ket{0}$, where $U$ and $W$ might be realized by sequences of parametrized and fixed quantum gates. Let $Y(x) = \bra{0} U^\dagger S(x)^\dagger W^\dagger H WS(x)U\ket{0}$ be the natural objective of the QNN with a Hermitian observable $H$. One can assume the observable and the encoding block are diagonal, i.e. $H = \text{diag}(h_1,\dots,h_d), h_j \in \mathbb{R}$ and $S(x) = \text{diag}(e^{i\lambda_1},\dots,e^{i\lambda_d})$. The function value can be written as a partial Fourier sum

\begin{equation}
\begin{aligned}
    Y &= \sum_{ijk=1}^d (U^\dagger)_{1i} (S^\dagger)_{ii} (W^\dagger)_{ij} H_{jj} W_{jk} S_{kk} U_{k1} \\
    &= \sum_{ijk} h_j e^{i(\lambda_k - \lambda_i)} \bar u_{i1}u_{k1} \bar w_{ji} w_{jk} \\
    &=\sum_{ik} e^{i(\lambda_k - \lambda_i)} \bar u_{i1}u_{k1} \sum_{j} h_j \bar w_{ji} w_{jk} \\
    &= \sum_i |u_{i1}|^2 \sum_j h_j |w_{ji}|^2 + \sum_{i\neq k} e^{i(\lambda_k - \lambda_i)} \bar u_{i1}u_{k1} \sum_{j} h_j \bar w_{ji} w_{jk}
\end{aligned}
\end{equation}

Denote (1) $\alpha_i = |u_{i1}|^2$, (2) $a_i = \sum_{j} h_j |w_{ji}|^2$, (3) $\beta_{ik} = \bar u_{i1}u_{k1}$, (4) $b_{ik} = h_j \bar w_{ji}w_{jk}, i,k \in [d], i\neq k$. We show in Appendix \ref{appx:coeff-dist} that 
\begin{equation}
    \begin{aligned}
    \alpha_i &\sim \operatorname{Beta}(1,d-1) \\
    a_i &\sim \mathcal{N}\left(1, \frac{(d-1)}{(d+1)} \Tr(H^2)  \right) \\
    \Re \beta_{ik}, \Im \beta_{ik} &\sim \operatorname{Laplace}(0,1/2d) \\
    \Re b_{ik}, \Im b_{ik} &\sim \mathcal{N}\left(0, \frac{1}{2d^2} \Tr(H^2) \right) 
\end{aligned}
\end{equation}

We continue to write
\begin{equation}
    \begin{aligned}
    Y &= \sum_i \alpha_i a_i + \sum_{i \neq k} e^{i(\lambda_k - \lambda_i)} \beta_{ik} b_{ik} \\
    &= \sum_i \alpha_i a_i + 2 \sum_{i<k} \Re \left\{ e^{i(\lambda_k - \lambda_i) \beta_{ik} b_{ik}} \right\} \\
    &= \sum_i \alpha_i a_i  + 2 \sum_{i<k} \cos(\lambda_k - \lambda_i) (\Re\beta_{ik} \Re b_{ik} - \Im\beta_{ik} \Im b_{ik}) \\
    & \quad \quad \quad \quad \quad \quad - \sin(\lambda_k - \lambda_i) (\Re\beta_{ik} \Im b_{ik} + \Im\beta_{ik} \Re b_{ik})
    \end{aligned}
\end{equation}

We again define real variables $\gamma_{ik} = \Re \beta_{ik}, \delta_{ik} = \Im \beta_{ik}$ and $c_{ik} = \Re b_{ik}, d_{ik} = \Re b_{ik}$. Then 
\begin{equation}
    Y = \sum_i \alpha_i a_i + 2\sum_{i<k} \cos(\lambda_k - \lambda_i)(\gamma_{ik} c_{ik} - \delta_{ik} d_{ik}) - \sin(\lambda_k - \lambda_i)(\gamma_{ik} d_{ik} + \delta_{ik} c_{ik})
\end{equation}
Since $U$ and $W$ each has $d^2$ real degree of freedom, we have the full control of $U$ via $d^2$ real variables $\alpha_i, \gamma_{ik}, \delta_{ik}$. This suggests us to use $\alpha_i, \gamma_{ik}, \delta_{ik}$ as independent parameters of $U$. It is similar for $W$ and $b_i, c_{ik}, d_{ik}$. The gradient of the function with respect to those variables is given by
\begin{equation}
\begin{aligned}
    \partial_{\alpha_i} Y &= a_i \\
    \partial_{a_i} Y &= \alpha_i \\
    \partial_{\gamma_{ik}} Y &= 2 \cos(\lambda_k - \lambda_i)c_{ik} - 2\sin(\lambda_k - \lambda_i)d_{ik} \\
    \partial_{\delta_{ik}} Y &= -2 \cos(\lambda_k - \lambda_i)d_{ik} - 2\sin(\lambda_k - \lambda_i)c_{ik} \\
    \partial_{c_{ik}} Y &= 2 \cos(\lambda_k - \lambda_i)\gamma_{ik} - 2\sin(\lambda_k - \lambda_i)\delta_{ik} \\
    \partial_{d_{ik}} Y &= -2 \cos(\lambda_k - \lambda_i)\delta_{ik} - 2\sin(\lambda_k - \lambda_i)\gamma_{ik}
\end{aligned}
\end{equation}
The tangent kernel function $K(x,x') = \nabla Y^T(x) \nabla Y(x')$ is equal to
\begin{equation}
\begin{aligned}
    K(x,x') &= \sum_i a_i^2 + \sum_i \alpha_i^2 \\
    &+ 8\sum_{i<k} \cos(\lambda_k - \lambda_i) \cos(\lambda_k' - \lambda_i')(c^2_{ik} + \gamma^2_{ik}) \\
    &+8\sum_{i<k} \sin(\lambda_k - \lambda_i) \sin(\lambda_k' - \lambda_i')(d^2_{ik} + \delta^2_{ik}) 
\end{aligned}
\end{equation}
This form of kernel function might be shown to be universal~\cite{schuld2021effect} given as many qubits as needed. It is simple to verify the following expected value: $\mathbb{E}[\alpha_i^2] = \frac{2}{d(d+1)}$, $\mathbb{E}[a_i^2] = 1+\frac{d-1}{d+1}\Tr(H^2)$, $\mathbb{E}[\gamma_i^2] = \mathbb{E}[\delta_i^2] = \frac{1}{2d^2}$, and $\mathbb{E}[c_i^2] = \mathbb{E}[d_i^2] = \frac{1}{2d^2}\Tr(H^2)$.

The analytical limit of the tangent kernel at the infinite limit is
\begin{equation}
    \begin{aligned}
    \mathbb{E}[K(x,x')] &= \frac{2}{d+1} + \left(d+\frac{d(d-1)}{d+1}\Tr(H^2)\right) \\
    & + \frac{4}{d^2}(1+\Tr(H^2)) \sum_{i<k} \cos(\lambda_k - \lambda_i) \cos(\lambda_k' - \lambda_i') \\
    & + \frac{4}{d^2}(1+\Tr(H^2)) \sum_{i<k} \sin(\lambda_k - \lambda_i) \sin(\lambda_k' - \lambda_i') \\
    &= d + \frac{2+d(d-1)\Tr(H^2)}{d+1} \\
    & + \frac{4}{d^2}(1+\Tr(H^2)) \sum_{i<k} \cos(\lambda_k - \lambda_i - \lambda_k' + \lambda_i').
\end{aligned}
\end{equation}
One might proceed to use this expected kernel for regression with kernel method. For robustness, it is recommended to normalize $Y$ by $\frac{1}{\sqrt{d}}$ to avoid $\Theta(d)$ terms in the expression of $\mathbb{E}[K(x,x')]$.

\section{Conclusion}
In this manuscript we have show the convergence of quantum neural tangent kernel at large dimension limit. For both Quantum Ensemble and Quantum Neural Network models, the limit of the kernel function has a concise analytical form from which efficient evaluation or estimation could be carried out. We expect this could aid the finding of new quantum models as an effective benchmark for the performance of supervised learning problems.

\printbibliography

\appendix

\section{Probability distribution of expected value of Hermitian observables}
\label{appx: observable-dist}
We want to determine the distribution of $Y = Y(\ket{\psi}) = \bra{\psi} H \ket{\psi}$ when $\ket{\psi}$ is sampled with respect to the Haar measure of quantum states, i.e. finding pdf $p(y)$.

To begin with, we parametrize the complex projective space $CP^{N-1}$ almost everywhere with the coordinates $\xi = [\xi_1,\dots,\xi_{N-1}]^T \in \mathbb{C}^{N-1}$. Express a random $N$-dimensional normalized vector corresponding to the point $\xi$ using local coordinates
\begin{equation}
    \ket{\phi(\xi,\bar\xi)} = \frac{[1,\xi_1,\dots,\xi_{N-1}]^T}{\sqrt{1+\xi^\dagger \xi}}
\end{equation}
The Fubini Study volume element is given by
\begin{equation}
    d\mu_{CP^{N-1}} (\xi,\bar\xi) = \frac{(N-1)!}{\pi^{N-1}} \frac{\prod_{k=1}^{N-1} d\xi_k d\bar\xi_k}{(1+\xi^\dagger \xi)^N}.
\end{equation}

In general, assume $H$ can be diagonalized into $\text{diag}(h_0, \dots, h_{N-1})$ in which the eigenvalues are in descending order. The pdf is given by:
\begin{align}
    \nonumber p(y) &= \int_{CP^{N-1}} \delta \left( \frac{h_0 + \sum_{k=1}^{N-1} h_k |\xi_k|^2}{1+\xi^\dagger \xi} - y \right) \frac{(N-1)!}{\pi^{N-1}} d\mu_{CP^{N-1}}(\xi, \bar\xi) \\
    &= \int_{CP^{N-1}} \delta \left( \frac{h_0 + \sum_{k=1}^{N-1} h_k |\xi_k|^2}{1+\xi^\dagger \xi} - y \right) \frac{(N-1)!}{\pi^{N-1}} \frac{\prod_{k=1}^{N-1} d\xi_k d\bar\xi_k}{(1+\xi^\dagger \xi)^N}
\end{align}

Since $|\xi_k|^2 = \| [\Re{\xi_k}, \Im{\xi_j}] \|$, we can replace $\xi_k$ by two real variables that are its real and imaginary parts. Apply a change of variable
\begin{equation}
    \Re \xi_k = r_k\cos(\theta_k), \Im\xi_k = r_j\sin(\theta_k),
\end{equation}
with
\begin{equation}
    \left| \frac{\partial(\Re \xi_k, \Im \xi_k)}{\partial (r_k, \theta_k)}\right| = r_k
\end{equation}
followed by another change of variable $z_k = r^2_k, k=1,\dots,N-1$. The pdf becomes
\begin{align}
    \nonumber &= \frac{(N-1)!}{2^{N-1}} \int_{r_1,\dots,r_{N-1}=0}^\infty \delta \left(\frac{h_0 + \sum_{k=1}^{N-1} h_k r^2_{k}}{1+ \sum_{k=1}^{N-1}r_k^2} - y \right) \frac{r_1 \dots r_{N-1}}{(1+ \sum_{k=1}^{N-1}r_k^2)^N} \prod_{k=1}^{N-1} dr_k \\
    &= (N-1)! \int_{z_1,\dots,z_{N-1}=0}^\infty \delta \left(\frac{h_0 + \sum_{k=1}^{N-1} h_k z_{k}}{1+ \sum_{k=1}^{N-1}z_k} - y \right) \frac{\prod_{k=1}^{N-1} dz_k}{(1+ \sum_{k=1}^{N-1}z_k)^N}
\end{align}
We want to get rid of the Dirac Delta function. Consider
\begin{equation}
    g(z_{N-1}) = \frac{h_0 + \sum_{k=1}^{N-2} h_kz_k + h_{N-1}z_{N-1}}{1 + \sum_{k=1}^{N-2} z_k + z_{N-1}}
\end{equation}
Its derivative is
\begin{equation}
    g'(z_{N-1}) = \frac{h_{N-1} - h_0 + \sum_{k=1}^{N-2} (h_{N-1} - h_k)z_k}{(1 + \sum_{k=1}^{N-2} z_k + z_{N-1})^2}
\end{equation}
Due to the order of $h_k$, the derivative $g'(z_{N-1})$ is negative, i.e. $|g'(z_{N-1})| = -g'(z_{N-1})$. Also, the peak of the Dirac Delta function occurs at 
\begin{align}
    z^*_{N-1} &= \frac{y-h_0 + \sum_{k=1}^{N-2}(y-h_k)z_k}{h_{N-1}-y} \\
    1+\sum_{k=1}^{N-2}z_k + z^*_{N-1} &= \frac{h_{N-1}-h_0 + \sum_{k=1}^{N-2}(h_{N-1} - y +1-h_k}{h_{N-1} - y} \\
    \Rightarrow g'(z^*_{N-1}) &=  \frac{(h_{N-1}-y)^2 (h_{N-1}-h_0 + \sum_{k=1}^{N-2} z_k)}{\left( h_{N-1}-h_0 + \sum_{k=1}^{N-2} (h_{N-1} - y + 1 - h_k)z_k \right)^2}
\end{align}
Therefore we rewrite
\begin{equation}
\begin{aligned}
    \delta \left(\frac{h_0 + \sum_{k=1}^{N-1} h_k z_{k}}{1+ \sum_{k=1}^{N-1}z_k} - y \right) &= \frac{\delta (z_{N-1} - \frac{y-h_0 + \sum_{k=1}^{N-2}(y-h_k)z_k}{h_{N-1}-y})}{|g'(z^*_{N-1})|} \\
    &= \delta \left(z_{N-1} - \frac{y-h_0 + \sum_{k=1}^{N-2}(y-h_k)z_k}{h_{N-1}-y} \right) \\
    & \times \frac{\left( h_{N-1}-h_0 + \sum_{k=1}^{N-2} (h_{N-1} - h_k + 1-y)z_k \right)^2}{(h_{N-1}-y)^2 \left(h_{N-1}-h_0 + \sum_{k=1}^{N-2} z_k\right)}.
\end{aligned}
\end{equation}
This allows us to eliminate the Dirac Delta function upon integrating over $z_{N-1}$. In addition since $g(0) > g(\infty)$, the density $p(y)$ is positive only when $g(0) \geq y > g(\infty)$, i.e.
\begin{align}
    h_{N-1} < y &\leq \frac{h_0 + \sum_{k=1}^{N-2} h_kz_k }{1 + \sum_{k=1}^{N-2} z_k}
\end{align}
The first constraint is trivial. We use the second constraint to derive bounds for $z_k$. It is equivalent to
\begin{equation}
    h_0 - y + \sum_{k=1}^{n-2}(h_k-y) z_k \geq 0
\end{equation}
Note that $p(y) = 0$ for $y \geq h_0$. Assume $h_{m-1} > y > h_m$ for some integer $1 \leq m \leq N-1$. Rewrite the constraint as
\begin{align}
    \sum_{k=1}^{N-2} (y - h_k)z_k \leq h_0 - y + \sum_{k=1}^{m-1} (h_k-y)z_k
\end{align}
Notice $h_k - y > 0, k=0,\dots,m-1$ and $y-h_k > 0, k=m,\dots,N-2$. Thus the bounds for $z_1,\dots, z_{N-2}$ are given as follows
\begin{align*}
            &\text{from} \quad& \quad \text{to} \\
    z_{N-2} \quad & 0 & \frac{h_0 - y + \sum_{k=1}^{m-1} (h_k-y)z_k + \sum_{k=m}^{N-3}(h_k-y)z_k}{y-h_{N-2}} \\
    z_{N-3} \quad & 0 & \frac{h_0 - y + \sum_{k=1}^{m-1} (h_k-y)z_k + \sum_{k=m}^{N-4}(h_k-y)z_k}{y-h_{N-3}} \\
    \dots \\
    z_{m} \quad & 0 & \frac{h_0 - y + \sum_{k=1}^{m-1} (h_k-y)z_k}{y-h_m} \\
    z_{m-1},\dots,z_2,z_1 \quad & 0 \quad & \infty
\end{align*}
The general procedure is follows. For each of non-trivial gaps $y \in (y_{m-1},y_m)$, we derive the bounds of variables $z_1,\dots,z_{N-1}$ as above, which can be used to compute, say by a computer-algebra system (CAS),
\begin{equation}
    p(y) = \frac{(N-1)!}{(h_{N-1}-y)^2} \int \frac{\left( h_{N-1}-h_0 + \sum_{k=1}^{N-2} (h_{N-1} - h_k + 1-y)z_k \right)^2}{h_{N-1}-h_0 + \sum_{k=1}^{N-2} z_k} dz_{N-2}\dots dz_1
\end{equation}

\section{Analytical form for Pauli-$Z$ encoding}
\label{appx: pauliZ-encoding}
Assume $S(x) = (e^{-ixZ})^{\otimes n} = \operatorname{diag}(e^{-ix}, e^{ix})^{\otimes n}$. Let $\Vec{v}(k) \in \{0,1\}^n$ be the binary representation of position index $k, k \in \{0,\dots,2^n-1\}$ along the diagonal matrix. The number of ``1'' and ``0'' in $v$ are $\expval{\Vec{1},\Vec{v}}$ and $\expval{\Vec{0},\Vec{v}}$ respectively, where $\Vec{1}$ is the all-one vector. One can verify that the expansion of $S(x)$ is a diagonal matrix whose $k$-th element is
\begin{equation}
    \begin{aligned}
    e^{ix\expval{\Vec{1},\Vec{v}(k)} - \expval{\Vec{1},\Vec{1}-\Vec{v}(k)}} &= e^{ix\expval{\Vec{1},2\Vec{v}(k) - \Vec{1}}} \\
    & \equiv e^{ixq_k}
\end{aligned}
\end{equation}
We can write
\begin{equation}
\begin{aligned}
\Tr(S_2S_1^\dagger)\Tr(S_2^\dagger S_1) &= \sum_{j,k} e^{i(x_2-x_1)(q_j - q_k)} \\
&= \sum_{j,k} \cos[(x_2-x_1)(q_j - q_k)
]
\end{aligned}
\end{equation}
Observe that $q_k$ can take any value in the set $\{-n, -n+2, \dots, n-2, n\}$ with $\binom{n}{0}, \binom{n}{1}, \dots, \binom{n}{n-1}, \binom{n}{n}$ occurrences among $2^n$ values of $k$. Therefore the value $q_j - q_k$ is belongs the set $\{0, \pm 2, \pm 4, \dots, \pm 2n\}$. We count the occurrences of those values. There are $\sum_{k=0} \binom{n}{k} \binom{n}{k} = \sum_{k=0} \binom{n-k}{k} \binom{n}{k} = \binom{2n}{n}$ combinations of $j$ and $k$ that leads to $q_j - q_k = 0$. Similarly, the occurrences of $\pm 2, \pm 4,\dots, \pm 2n$ are $\binom{2n}{n-1}, \binom{2n}{n-2}, \binom{2n}{0}$ respectively, i.e.
\begin{equation}
    \left |\Tr(S_2S_1^\dagger) \right|^2 = \binom{2n}{n} + 2\sum_{k=1}^n \binom{2n}{n-k} \cos(2k(x_1-x_2))
\end{equation}

\section{Distribution of coefficients}
\label{appx:coeff-dist}
It is known that each entry from a Haar random unitary is $\operatorname{Beta}(1,d-1)$ distributed, i.e. $P(|u_{ij}|^2 = x) = (d-1)(1-x)^{d-2}$ \cite{zyczkowski2005average}, which can also be derived following the technique in Appendix \ref{appx: observable-dist}. The product $h_j |w_{ji}|^2$ therefore has the mean of $\frac{h_j}{d}$ and the variance of $\frac{h_j^2(d-1)}{d^2(d+1)}$. Moreover these random variables are almost surely independent. Because $\frac{1}{d}\sum_{j} \operatorname{Var}(h_j |w_{ji}|^2) = \sum_{j} \frac{h_j^2(d-1)}{d(d+1)}$ converges to $0$ when $h_j \in O(1)$, the Lindeberg-Feller Central Limit Theorem \cite{kallenberg1997foundations} (Theorem 6.13) says that the sum $\sum_j h_j |w_{ji}|^2$ converges in distribution to a normal random variable $\mathcal{N}\left(1, \frac{(d-1)}{(d+1)} \sum_{j} h_j^2  \right)$. 

On the other hand, we expand $\bar u_{i1} u_{k1}$ with their real and imaginary parts.
\begin{equation}
\begin{aligned}
    \bar u_{i1} u_{k1} &= (x_{i1} - iy_{i1})(x_{k1} + iy_{k1}) \\
    &= (x_{i1}x_{k1} + y_{i1}y_{k1}) + i(x_{i1}y_{k1} - x_{k1}y_{i1})
\end{aligned}
\end{equation}
We know that the real and imaginary part of any entry of a Haar random unitary are identically independently distributed from $\mathcal{N}\left(0, \frac{1}{2d}\right)$ with the corresponding characteristic function $\varphi(t) = e^{-t^2/4d}$. Moreover, the components of $u_{i1}$ are almost surely independent to the components of $u_{k1}$. The characteristic function of $x_{i1}x_{k1}$ is therefore
\begin{equation}
\begin{aligned}
    \mathbb{E}[\exp(it x_{i1}x_{k1})] &= \mathbb{E}[\mathbb{E}[\exp(it x_{i1}x_{k1}) | x_{k1}]] \\
    &= \mathbb{E} [\exp(-t^2 x_{k1}^2/4d)] \\
    &= \int_{-\infty}^\infty \frac{\exp(-x_{k1}^2 d)}{\sqrt{\pi/d}}\exp(-t^2 x_{k1}^2/4d) dx_{k1}
\end{aligned}
\end{equation}
The last line is a Gaussian integral with $\sigma^2 = \frac{2d}{4d^2+t^2}$, hence evaluated to be $1/\sqrt{1+(t/2d)^2}$. Thanks to the independence, the characteristic function of $x_{i1}x_{k1} + y_{i1}y_{k1}$ is $1/(1+(t/2d)^2)$, equal to the characteristic function of the distribution $\operatorname{Laplace}(0,1/2d)$. The same argument applies for $x_{i1}y_{k1} - x_{k1}y_{i1}$. 

The previous argument also reveals that $\sum_j h_j \bar w_{ji}w_{jk}$ has both real and imaginary parts being two linear combinations of (almost surely independent) random variables with distribution from $\operatorname{Laplace}(0,1/2d)$. In particular, both real and imaginary parts have the form of $\sum_j q_j$, where $q_j \sim \operatorname{Laplace}(0,|h_j|/2d)$ with $\operatorname{Var}(q_j) = h_j^2/2d^2$. Since $\frac{1}{d} \sum_j \frac{h_j^2}{2d^2} = \frac{\sum_j h_j^2}{2d^3} \rightarrow 0$ when $h_j \in O(1)$, the Lindeberg-Feller Central Limit Theorem concludes $\Re \left \{\sum_j h_j \bar w_{ji}w_{jk} \right\}$ and $\Im \left \{\sum_j h_j \bar w_{ji}w_{jk} \right\} \xrightarrow[]{\text{dist.}} \mathcal{N}\left(0, \frac{1}{2d^2}\sum_j h_j^2\right)$.
\end{document}